# Towards a unified approach to formal "risk of bias" assessments for causal and descriptive inference

O.L. Pescott[a*], R.J. Boyd[a], G.D. Powney[a] & G.B. Stewart[b]
a. UK Centre for Ecology & Hydrology, Wallingford, UK, OX10 8BB
b. Evidence Synthesis Lab, School of Natural and Environmental Science, University of Newcastle, Newcastle-upon-Tyne, NE1 7RU
*Corresponding author: olipes@ceh.ac.uk

## Abstract

Statistics is sometimes described as the science of reasoning under uncertainty. Statistical models provide one view of this uncertainty, but what is frequently neglected is the "invisible" portion of uncertainty: that assumed not to exist once a model has been fitted to some data. Systematic errors, i.e. bias, in data relative to some model and inferential goal can seriously undermine research conclusions, and qualitative and quantitative techniques have been created across several disciplines to quantify and generally appraise such potential biases. Perhaps best known are so-called "risk of bias" assessment instruments used to investigate the likely quality of randomised controlled trials in medical research. However, the logic of assessing the risks caused by various types of systematic error to statistical arguments applies far more widely. This logic applies even when statistical adjustment strategies for potential biases are used, as these frequently make assumptions (e.g. data "missing at random") that can rarely be empirically guaranteed. Mounting concern about such situations can be seen in the increasing calls for greater consideration of biases caused by nonprobability sampling in descriptive inference (e.g. in survey sampling), and the statistical generalisability of in-sample causal effect estimates in causal inference. Both of these relate to the consideration of model-based *and* wider uncertainty when presenting research conclusions from models. Given that model-based adjustments are never perfect, we argue that qualitative risk of bias reporting frameworks for both descriptive and causal inferential arguments should be further developed and made mandatory by journals and funders. It is only through clear statements of the limits to statistical arguments that consumers of research can fully judge their value for any given application.

### Keywords

Internal validity; external validity; generalisability; transportability; risk-of-bias; survey sampling; experiments; qualitative methods; estimand

## Introduction

At its most general, bias is normally defined as a systematic deviation from the truth (Delgado-Rodríguez & Llorca, 2004). This definition, with its link to everyday notions of balance and even-handedness when dealing with the world, underlies the more technical usages that we encounter as researchers, such as considerations of biased and unbiased estimators in frequentist statistics (Wasserman, 2004). Other technical definitions of specific biases abound in research (e.g. see the Catalogue of Bias, https://catalogofbias.org, originally proposed by Sackett, 1979), although these often relate to extra-statistical issues that can undermine the use of "unbiased" or any other approaches to statistical estimation (Greenland & Pearce, 2015). Perhaps the best known examples of these are selection bias in survey sampling (i.e. some non-zero correlation between the variable of interest in the population and the probability of being selected into a sample; Meng, 2018), and practical problems in the execution of randomised controlled trials (e.g. subject drop-out) resulting in the confounding of a potential treatment effect with other covariates (Higgins et al., 2019).

Different statistical approaches to modelling the world consider aspects of bias differently. For example, Bayesian statistics does not contain the idea of unbiased estimation of parameters in the same way as frequentist statistics: unknown parameters in a Bayesian model require prior distributions, and in this framework a posterior distribution for a parameter cannot be biased in the classical asymptotic sense, as it is the direct product of the prior and the joint model specified for the parameters and data (Gelman et al., 2014). Here discussions of bias require either some external reference point or a frequency-based evaluation of Bayesian estimates (Gelman et al., 2014). Even within a classical framework biased estimators are not completely outlawed; they can be acceptable if the bias in an estimator is outweighed by a reduction in its variance (Gelman, 2008; Hodges, 2014). This idea underlies the widespread "empirical Bayes" approach to the use of random effects in multilevel models (Efron & Hastie, 2016; Snijders & Bosker, 2012), where group-level estimates with low precision are shrunk towards a group mean, with the aim of improving predictive inferences on average (Efron & Morris, 1975).

The potential utility of this bias–variance trade-off, or indeed the choice between frequentist and Bayesian philosophies (Gelman et al., 2021 p. 16) or any set of models, highlights the importance of what Hodges (1996) called the statistical "argument" of an analysis: the practical value of any given modelling exercise can only be evaluated relative to a real-world inferential goal (Greenland, 2021). For example, partial pooling (i.e. shrinkage, or, with added implicit value judgment, "borrowing strength") may be desirable where some groups are more uncertain than others, and we hope to make a minimum-error estimate or prediction. In other cases, where an analyst is tasked with directly describing populations with their attendant uncertainty, for example in a descriptive social survey or the production of an indicator (*sensu* Gerring, 2012), "no pooling" approaches may be preferable. Similarly, defensible arguments may be put forward for different analytical approaches to the same real-world problem (Draper, 1995; Gelman & Hennig, 2017; Steegen et al., 2016). Statistical models should therefore always be understood and presented in the context of their underlying argument (Hodges, 1996). This logically extends to unmodelled sources of plausible bias, because the omitted parts of the "real world" may influence model-based conclusions just as much as those in the specified model structure (Gelman et al., 2021; Greenland, 1998; Greenland & Pearce, 2015; Lash et al., 2009, 2014; Saltelli et al., 2020; Stark, 2022). These considerations motivate treating bias as a property of the inferential claim, not merely of an estimator or design. We therefore begin by framing applied statistics as argumentation and by making the relevant "burdens of proof" explicit across inferential tasks. Figure 1 previews the central distinction used throughout, between identification within the study setting and the separate step of asserting a claimed target of inference, while Box 1 provides a minimal worked illustration in both causal and descriptive settings.

## Statistical arguments in scientific research

As far as we are aware there is as yet no single accepted theory or classification of statistical arguments. Mallows (1998) argued for the fundamental importance of such a framework, adding the "zeroth problem" of statistics, namely consideration of the relevance of data in hand (or which might be observed) to a substantive problem, to Fisher's (1922) canonical trio of model specification, estimation given a model and the sampling distributions of estimators. Hodges (1996) attempted a classification of the arguments implied by different statistical approaches, identifying five logical types: (i) causal arguments; (ii) non-causal predictive arguments; (iii) description arguments; (iv) existence arguments; and (v), hypothesis-generation arguments. Each of these may have action and non-action subtypes, thereby also covering decision-theoretic aspects of statistics. Hodges' argument for this framework was that it helped to identify the "burdens of proof" against which models could be evaluated. Mallows (1998) noted how this highlighted that statistical arguments were not just the product of a model and some calculations, but "[included] the assumptions, the logical steps involved, and the conclusions."

The first three of Hodges' (1996) types have often been distinguished by other researchers: Shmueli (2010) identified explanatory, predictive and descriptive aims for models, whilst Hernán et al. (2019) similarly distinguished between description, prediction and causal inference (see Carlin & Moreno-Betancur, 2025 for a recent review of this classification). Others have divided the territory differently, and used different language, but have arrived at similar mappings. For example, Gelman et al. (2021) list "generalization from sample to population, and from treatment to control group" as two of their three goals of statistics. Depending on the desired inferences, it is not difficult to see how these aims can be mapped onto the "explain, predict, describe" triumvirate. Our aim here is neither to choose a canonical set, nor to pick over small differences in outlook, but to highlight the broad agreement between researchers that different statistical aims, including causation and description, are normally distinguished.

Different logical frameworks will be required for different inferential aims, and so it follows that specific model/inference combinations will have particular burdens of proof (Hodges, 1996; Kenett & Shmueli, 2014). For causal arguments, the burden is often "to rule out all but one causal agent" (Hodges, 1996); practically, this may mean that causal interpretations of coefficients in a model are not likely to be justified if no thought has been given to linkages between variables in the real world (Cinelli et al., 2022). For descriptive arguments, the burden of proof is "showing that the inferential statement has the properties claimed for it" (Hodges, 1996). For example, has the sample in-hand arisen from a probabilistic mechanism of some sort, and so can it be claimed to be representative of some intended target population (Figure 1)? Whilst these are different types of statistical problem, the required evidence for accepting statistical conclusions in each case is not merely statistical. At a minimum extra-statistical causal logic will be required in the first case; in the second, information about the sampling mechanism, or properties of the sample, may be required as soon as some departure from probability sampling is encountered, or if the sample is small (Meng, 2018). Given these points, it is instructive to consider the similarities and differences between descriptive and causal inferential aims, how they attempt to deal with threats to their validity, and to examine how each might learn from the other (Hedlin, 2015; Mercer et al., 2017)[1]. The next section summarises the shared structure of these inferential problems and the consequences for assessing potential biases.

## The many links between descriptive and causal inference

It has long been noted that the links between experimental design and survey sampling run deep through statistics (Fienberg & Tanur, 1987, 2018; Hedlin, 2015): both logics rely on randomisation for their strongest inferences, either through probability sampling from a population (description), or through random treatment assignment (causal attribution) (Greenland, 1990). Both areas also contain long-standing debates relating to the use of design probabilities (the design-based paradigm in survey sampling and the randomisation–inference paradigm in causal inference) or model-based probabilities for inference (Ding et al., 2017; Meng, 2022; Valliant et al., 2000).

Attempts in both areas to clarify the situations in which estimates of various different estimands may be biased has led to much exchange of ideas. For example, Rubin's (1976) missing data framework has been used to clarify biases resulting from missingness in both causal settings (e.g. Howe et al., 2015; Smith & Sugden, 1988) and nonprobability survey samples (Mercer et al., 2017; Smith, 1983; Sugden & Smith, 1984). In survey sampling the logic has also recently helped to clarify the modes of action of different bias adjustment methods (Bailey, 2023; Chen et al., 2020; Meng, 2022; Wu, 2022). Causal graphs (i.e. network diagrams encoding assumptions about links between variables) have been used in both settings to further elucidate the logical basis, and impacts, of adjusting for auxiliary variables[2] in missing data situations (Mohan & Pearl, 2021; Thoemmes & Mohan, 2015). Consideration of the precise consequences of adjustment variables for specific estimands can be combined with causal graphs for missingness in order to move beyond the general advice to adjust for as many biasing variables as possible in both causal inference and survey sampling (Boyd et al., 2025; Cinelli et al., 2022).

Quantitative solutions to departures from probability sampling in surveys tend to mirror the solutions put forward by causal inference practitioners for dealing with lack of covariate balance in experimentation and quasi-experimentation. For example, propensity scores in causal inference (Rosenbaum & Rubin, 1983, 2023) and the quasi-randomisation approach to nonprobability samples in survey sampling (Elliott & Valliant, 2017; Meng, 2022) can be seen as analogous approaches to reweighting units in order to approach covariate balance or population representation (Li et al., 2013). Survey sampling practitioners are also beginning to embrace approaches developed within causal inference that use explicit models for the correlation between sample inclusion and the response of interest. For example, Bailey (2023) outlines how the selection model of Heckman (1979), well-known in econometrics, could potentially improve practice in polling by encouraging researchers to investigate and consider models for non-ignorable non-response (i.e. data "not missing at random" *sensu* Rubin, 1976), rather than assuming that standard approaches to weighting[3] always produce reasonable answers that cannot be improved upon. This is not to say that the Heckman selection model will always be preferable for estimating population quantities in survey sampling, but the possession of a richer theoretical toolbox opens up new possibilities for modelling aspects of data that were previously stuck in the category of unmodelled error or ignored (Bailey, 2023).

---

[1] Note that we do not consider purely predictive inferences further here, although it seems clear that approaches that focus on appraising biases have much to offer in this domain too (e.g. see Wolff et al., 2019).

[2] That is, variables whose inclusion in a model can potentially help adjust for missing values in order to remove bias due to this (e.g. Collins et al., 2001).

An additional, and very direct, link between the causal and descriptive worlds is the statistical generalisability of a treatment effect estimate (Figure 1; Hartman, 2021; Lash & Rothman, 2021). (Here we use *generalisability* for extension from a study setting (e.g. study frame and/or realised sample) to its target population and *transportability* for out-of-frame/source population extension to a different target; Lash & Rothman, 2021.) Once a causal effect has been estimated within a sample, this can be considered as a description of a quantity in that sample that may or may not take the same value in some wider population. For example, approaches to reweighting the Sample Average Treatment Effect to estimate a Population Average Treatment Effect have relied on post-stratification (Hartman, 2021; Hedges & O'Muircheartaigh, 2010), in the same way that this has been a long-accepted method for reweighting nonprobability samples in survey sampling (Elliott & Valliant, 2017; Meng, 2022). Deciding on the appropriate effect moderators by which to stratify is analogous to deciding on appropriate auxiliary variables for the same purpose in survey sampling (Bailey, 2023; Hartman, 2021; Valliant et al., 2000).

The fact that the generalisability of a causal estimate is a very similar problem to survey-based descriptive inference yields numerous other tools that can be used in both situations. Many of these relate to evaluating the success of weighting (and related) techniques, rather than implementing adjustments. For example, recently proposed sensitivity analyses for survey weights are extensions of approaches developed for estimating causal effects with weighted estimators (Hartman & Huang, 2023). In the other direction, work on assessing the likely representativeness of samples relative to populations in survey sampling (e.g. Schouten et al., 2009) has led to proposals for indexing the likely generalisability of causal estimates (Tipton, 2014). Similar visualisation methods for assessing covariate or propensity score balance in causal inference (e.g. Aikens & Baiocchi, 2023; Greifer, 2023) and representativeness in survey sampling (e.g. Boyd et al., 2024; Makela et al., 2014) can also be found.

Given these numerous overlaps between the approaches that causal and descriptive inference practitioners use for dealing with real-life trials and observational quasi-experiments, and with survey samples that inevitably depart from asymptotic states of perfection (Meng, 2018), it is worth looking closely at practices on each side of the divide which have not so far made in-roads into the other's territory (Fienberg & Tanur, 2018; Hedlin, 2015). The shared logic of adjustment, ignorability, and generalisation suggests a shared vulnerability: systematic error may persist even after modelling choices are made. The next section reviews how existing "risk of bias" frameworks attempt to make the potential vulnerabilities in such claims explicit.

## "Risk of bias" thinking in primary research and evidence synthesis

Formal frameworks and tools for assessing any extra-statistical risk of bias are arguably most developed in health research. In this area "risk of bias" is normally taken as being synonymous with internal validity, although tools vary, and some researchers also consider aspects of external validity to be covered by this phrase as well (Wang et al., 2022). Well-known examples include the Cochrane RoB2 tool for randomised controlled trials (Sterne et al., 2019) and the ROBINS-I (Sterne et al., 2016) tool for observational studies. These assessment instruments are then, in turn, recommended within specific reporting guidelines: examples include PRISMA (Page et al., 2021) for systematic review and meta-analyses, STROBE (von Elm et al., 2007) for the reporting of observational epidemiological studies, and the GRADE framework for evaluating an evidence base in order to make recommendations for clinical practice (Guyatt et al., 2008).

[3] Such approaches sit within the "missing at random" category of Rubin's (1976) framework, in that they produce unbiased estimates in expectation if the value of the response conditional on some covariates is unrelated to the inclusion probability (Bailey, 2023; Wu, 2022).

In a recent review of the risk of bias tools available for randomised trials in medical research, Wang et al. (2022) argued for the internal validity definition of risk of bias, but, in the face of significant variation in tool design, concluded that:

*"[o]ne might reasonably argue that moving forward in this area requires establishing a consensus on classification of risk of bias vs. issues such as applicability/directness[4]. Doing so may involve coming to agreement regarding the concept of risk of bias, and differences between assessing risk of bias in the context of systematic reviews vs. single RCT. The problem with seeking such a consensus is that, for several such issues, alternative positions are reasonable, and possibly equally reasonable. In such situations, achieving consensus may not be a realistic or appropriate objective in the short term."*

Our, hopefully not unreasonable, position is that the simple phrase "risk of bias" contains no element that immediately signifies an exclusive concern with internal validity, and that for some statistical arguments (i.e. types of inference) the risks from external validity biases will be most important (e.g. Boyd et al., 2022). This is not to say that disciplines should not establish definitions that are useful for communication within their area, but, arguably, issues of statistical validity that bridge disciplinary and inferential divides may ultimately be better served by general usages that are at least not actively confusing. In short, risk of bias tools operationalise judgement about the credibility of a statistical argument, but their domain boundaries are often historically contingent rather than logically necessary. We now outline how an integrated risk of bias approach can be stated for both causal and descriptive inference.

## Towards a unified approach to risk of bias assessments across causal and descriptive inference

Regardless of whether and when random sampling and representation of any given population is required for useful experimental trials (e.g. see Rothman et al., 2013 and their discussants), it seems indisputable that extrapolation to new populations with characteristics that potentially modify a causal effect requires caution and additional work (Box 1; Hoggatt & Greenland, 2014). External validity biases in this case can only exist in relation to researcher claims and specific applications: as Shadish et al. (2002 p. 34, their emphasis) point out: "[v]alidity is a property of inferences. It is *not* a property of designs or methods, for the same design may contribute to more or less valid inferences under different circumstances". The results from a well-executed randomised controlled trial may be perfectly valid internally and externally for the group upon which the experiment was performed, but claiming that they are valid for a different population may result in bias (i.e. a systematic deviation from the truth)[5]. Once we acknowledge this, and begin to construct a logical framework for discussing the issue and proposing solutions (e.g. Degtiar & Rose, 2023; Findley et al., 2021; Hartman, 2021), then we are in very similar territory to problems which have been extensively discussed within descriptive inference (Lash & Rothman, 2021), and specifically that part of survey sampling dealing with nonprobability samples (Meng, 2022; Wu, 2022).

[4] Applicability is often used to denote whether or not the patients studied in a piece of research are representative of those of interest to the current clinical question. Indirectness has a broader circumscription, and covers this, as well as things like whether the treatment tested is relevant to the one of interest for the current clinical purpose (Guyatt et al., 2011).

[5] Whether or not this means that the phrase "external validity" properly differs from "generalisability" or not has been debated (Reichardt, 2011; Shadish, 2011), and it is important to remember that scientific generalisation (which, for example, Reichardt, 2011, links with external validity), is not the same as statistical generalisation, which is what many (if not most) researchers mean when they equate external validity with generalisability (e.g. Rothwell, 2005; Spake et al., 2022; Vazire et al., 2022). See Lash & Rothman (2021) for more discussion of this point.

The discussion of the generalisability (and transportability) of causal effects is the flip side of assessing the inferential risk of bias in its broadest sense. We are not the first to suggest this: one of the six recommendations of Franco et al. (2017) for reporting estimates of Population Average Treatment Effects using reweighted Sample Average Treatment Effect estimates was that "[i]f a sample does not vary on observables that plausibly moderate a treatment effect, such as when the sampling frame for a study excludes some demographic groups, researchers should discuss how this limits the generalizability of their findings and/or redefine their target population." That is, a qualitative assessment of the risk of bias to the inferential claims is recommended. Similarly, the "Constraints on Generality" (CoG) tool of Simons et al. (2017) within psychology suggests that researchers should clearly discuss limits to their experimental generalities based on comparisons between an experimental sample and the real world populations of which the researchers would like to speak (for similar ideas see Findley et al., 2021 and Degtiar & Rose, 2023). What has not apparently been widely suggested or adopted is that:

(i) the same checks should routinely be applied to descriptive inferences within survey sampling (or any other type of research that relies on accurate summaries of population states, e.g. Boyd et al., 2022, 2023; Jamieson et al., 2023); and,

(ii) formal, and ideally mandatory, tools (e.g. standardised assessment instruments or reporting guidelines) used for assessing internal validity biases in medicine and related fields should also be extended to issues of generalisability and transportability in other research areas in order to promote regular and clear contemplation of these issues. We doubt that general exhortations for improved practice (like this one) are likely to be sufficient (e.g. Sterba et al., 2011).

Our main aim in this paper has been to demonstrate that qualitative assessments of the "risk of bias" in causal research, taken to refer solely to internal validity biases in some areas (Wang et al., 2022), can be situated within a much wider framework that also covers the external validity of causal effects and survey sampling biases relevant to descriptive inferences concerning population quantities. The biases in these latter two categories are only present in relation to researcher claims (i.e. a formal statement of the estimand). A causal inference from a robust experimental or quasi-experimental framework may have high internal validity, but systematic errors in statements made off the back of statistical models can emerge in relation to wider claims, and these are just as much a part of the statistical argument as issues relating to model specification, estimation or internal validity (Greenland & Pearce, 2015; Hodges, 1996; Mallows, 1998). This framing clarifies why residual, unmodelled sources of systematic error remain relevant even in sophisticated modelling workflows (Pescott, 2023). We therefore close by stating what should be disclosed alongside results to allow readers to evaluate model fitness-for-purpose.

## Conclusions

As Gelman et al. (2021, p. 56) point out, "unmodelled error can be decisive in many real problems". In some cases part of the solution will be to expand the model, whether by improving a single model or expanding the universe of models considered (e.g. Draper, 1995; Greenland, 2005; Steegen et al., 2016). Unfortunately, in many cases important unmodelled error is likely to persist, and it is here that additional qualitative assessment and reporting may find its niche. Analysts have long pointed out the desirability of verbal caveats alongside modelled results (e.g. Cochran, 1983, pp. 13, 16, 73), and so to the many parallels between the causal work of experimentalists and quasi-experimentalists, and the typically descriptive work of survey samplers, we can and should add the creation of formal "risk of bias" assessments for the promotion of such considerations (e.g. Boyd et al., 2022).

In this, our recommendation is similar to those presented by others in recent years. For example[6], from a survey sampling perspective, Valliant (2023) argues for the presentation of "ignorability diagnostics" to be become "*de rigeur to justify the use* of any nonprobability sample for population estimation"; from data scientists, Meng (2021) requests "deeper data minding and *fuller data confessions*"; in political science Findley and colleagues (2021) have the "practical ambition […] that every published social science study *include a dedicated discussion of external validity*"; likewise, the statisticians Degtiar & Rose (2023) call for "*[clear descriptions of] the internal and external validity assumptions* needed to identify the treatment effect as they relate to the study"; whilst Rao (2021), another survey sampling statistician, points out that "[t]he evidence to date is that the use of nonresponse and calibration adjustments to compensate for nonresponse in probability samples can reduce nonresponse bias *but it will not eliminate it*", and so that the "dilemma for analysts of non-probability samples is *to assess how large is the residual nonresponse bias and whether the survey estimators are 'fit for purpose'*". We have already noted the call of Franco and colleagues (2017) for *standards in communicating weightings* for causal effect generalisation, and the "Constraints on Generality" initiative of Simons et al (2017) above. Sterba and colleagues (2011) go further, formalising these points as ultimately being an issue of research ethics, arguing that:

*"[t]here are several reasons why the gaps highlighted here between applied practice and methodological recommendations [concerning issues of sample representativeness] go beyond a purely methodological issue and into an ethical issue (e.g., negligence). These gaps are an ethical issue because researchers have the resources and ability to do something about them, but unintentionally have not, which leads to undesirable or even harmful consequences."*

Echoing all such concerns, we call for such assessments to be regularly codified into formal frameworks (e.g. Boyd et al., 2022) that are strongly encouraged (or, better, made mandatory) by funding agencies and publishers, and that they become an essential part of the scientific process where required by the statistical argument underlying a research claim (Hodges, 1996). At a minimum, authors should explicitly state the estimand and target population (e.g. Kahan & Devane, 2026), describe the study frame and/or realised sample (e.g. Jamieson et al., 2023), and summarise the assumed selection/nonresponse and missingness mechanisms and key measurement limitations that bear on generalisability or transportability.

We acknowledge that forcing form-filling cannot be unthinkingly equated with improvements in research quality, but a procedure through which data and modelled results must pass would at least force such information into the public domain, giving others some chance of assessing its adequacy (Boyd et al., 2023; Jamieson et al., 2023). We do not think that this approach is likely to be the only way of dealing with important unmodelled biases within disciplinary literatures, and complementary quantitative approaches to these problems will no doubt also become increasingly important for the transparent communication of uncertainty (e.g. Greenland, 2005; Hartman & Huang, 2023; Lash, 2007; Lash et al., 2009; Meng, 2022; Pescott et al., 2022; Steegen et al., 2016). The key attribute of all such systems is that they broaden the appreciation of consumers of statistics of additional sources of error not considered by the model(s) in hand (Blastland et al., 2020; Gelman et al., 2021; Greenland, 2021, 2023; van der Bles et al., 2019), avoiding what Manski (2011) memorably labelled "incredible certitude". Lash (2007) neatly frames the issue as one of converting an otherwise invisible part of uncertainty into the visible range. The mandatory use of risk of bias frameworks for internal validity in health research has encouraged this, and the joint attack of journal standards and requirements for evidence synthesis seem to us the main ways in which this may be realised in other disciplines (Van Calster et al., 2021). Indeed, as Van Calster and colleagues (2021) emphasise:

*"[t]op-down action from journals, funding agencies, universities and governments is needed to break the cycle. These actions should give methodology a central place in funding acquisition as well as study design, conduct, and reporting."*

---

[6] The italics are all ours in order to emphasise the strong language and clear agreement in this area across fields, theoreticians and practitioners.

More broadly, our philosophy is one of taking time for self-criticism (Berg & Seeber, 2016), and for taking the scientific challenge of falsifiability seriously (Mayo, 2018; Popper, 1959). Consideration of extra-statistical error is ultimately a core part of an analyst's duty (Gelman et al., 2021, p. 16), and an important issue for research ethics (Sterba et al., 2011), regardless of whether this relates to claims of experimental process, or to the statistical generalisability of in-sample estimates.

## Acknowledgements

OLP, RJB and GBS were supported by the NERC Exploring the Frontiers award number NE/X010384/1 "Biodiversity indicators from nonprobability samples: Interdisciplinary learning for science and society". OLP and GDP were also supported by the NERC award number NE/R016429/1 as part of the UK Status, Change and Projections of the Environment (UK-SCAPE) programme delivering National Capability. Any biases relative to the interests and perspectives of others are our own.

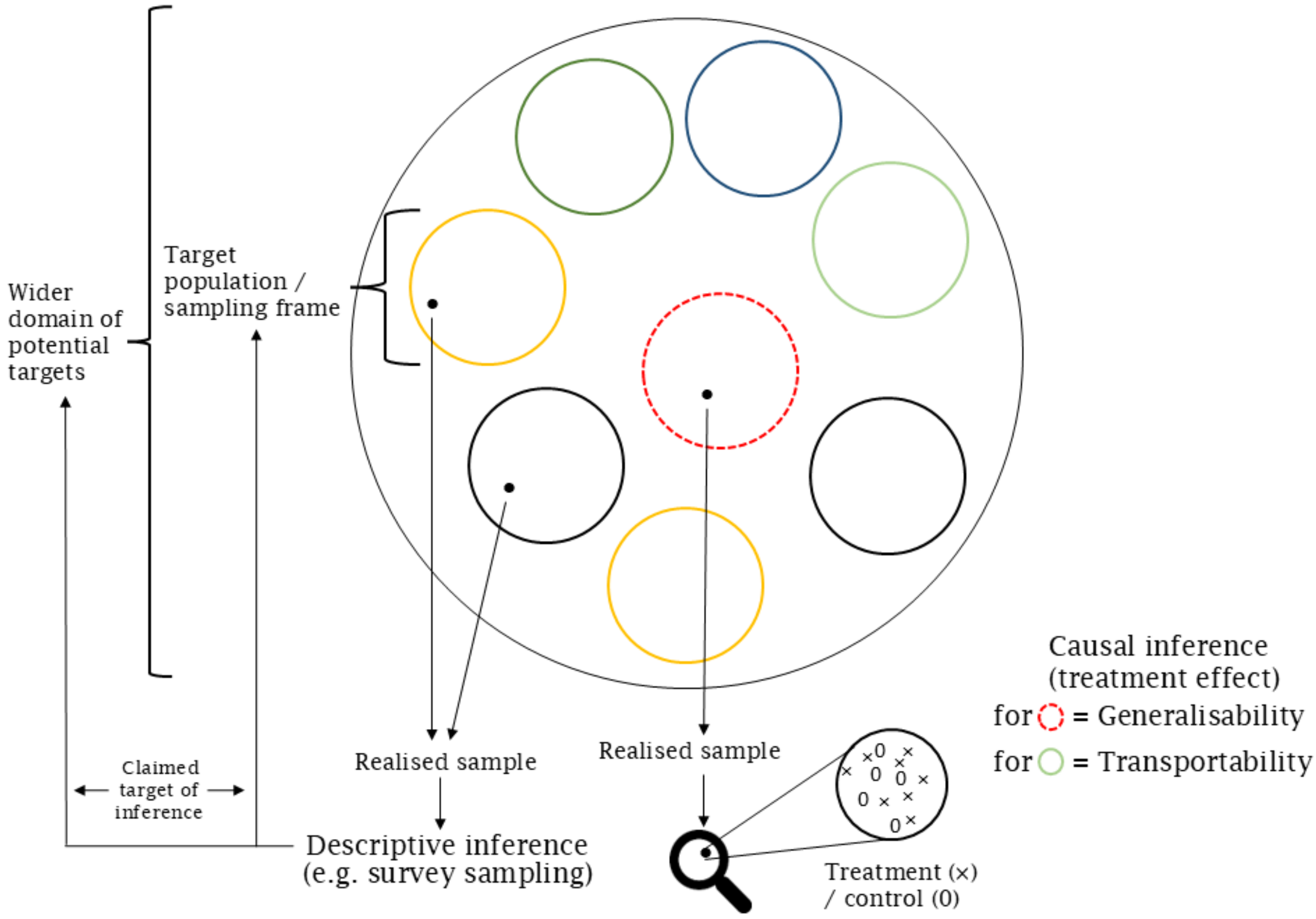

**Figure 1. Targets of inference and validity claims across descriptive and causal settings.** The study sample is drawn from a target population/sampling frame via a selection mechanism (note that in this example the population and sampling frame as assumed to coincide for simplicity). **Generalisability** refers to claims from the study sample to a target population, whereas **transportability** refers to extending the same estimand to a different target population. For causal questions, internal validity concerns identification of the causal estimand within the study setting; external validity concerns the additional assumptions required to apply that estimand to the claimed target population(s). In descriptive settings, validity is most transparently discussed in terms of the links between the realise sample *S*, any defensible source population (e.g. a sampling frame or eligibility-defined study population) *F*, and the stated target population *P*. One component concerns identification for the source population: whether selection/nonresponse, missingness and measurement processes permit unbiased (or plausibly low-bias) inference for the descriptive estimand defined on *F*. A second component concerns extension to the target: whether *F* adequately corresponds to *P*, or whether additional assumptions are required to justify out-of-frame generalisation/transport. For many nonprobability samples the source population *F* may be poorly defined, in which case these components cannot be cleanly separated and assumptions must be stated directly in terms of how *S* relates to *P* (Mooney & Garber, 2019).

**Box 1. A minimal worked illustration of the unified validity logic**

Consider two inferential claims that differ in estimands but share the same logical structure in Figure 1.

**(A) Causal claim (treatment effect)**

A randomised trial recruits from an eligible *target population/sampling frame* and yields a *realised sample*. The primary claim concerns a causal estimand (e.g., an average treatment effect) that is identified under internal validity conditions (e.g. randomisation, protocol adherence, outcome measurement, and missingness). A second, distinct step concerns the *claimed target of inference* beyond the study in question: **generalisability** requires that the realised sample and/or study frame are relevant to the target population, whereas **transportability** requires additional assumptions when the population for inference differs from the initial target population sampled, typically involving effect modification by covariates whose distribution changes across populations (Lash & Rothman, 2021).

**(B) Descriptive claim (population mean/prevalence/trend)**

A survey (or nonprobability sample) begins with a study frame and/or target population and produces a *realised sample*. Here the estimand is descriptive (e.g. a population mean, prevalence, or time trend). For descriptive claims, threats can arise (i) through coverage/frame correspondence (whether any defensible source population corresponds to the target), (ii) through selection/nonresponse step (whether the realised sample is representative of the source population, possibly after adjustment), and (iii) through measurement (whether observed variables correspond to the intended quantities). As noted in Figure 1, in some cases (e.g. some nonprobability samples), the source population may be poorly defined, in which case coverage and selection components cannot be cleanly separated and assumptions must be stated more directly in terms of how the realised sample relates to the stated target population.

**Interpretation**

Across both cases, the key is to separate (i) assumptions linking the realised sample to a defensible source population (selection/nonresponse, missingness, measurement) from (ii) assumptions required to extend from that source to the claimed target of inference (i.e. coverage and any out-of-frame extension/transport). For causal claims, these correspond to identification of the causal estimand within the study setting and to a subsequent generalisation/transport to the target population. For descriptive claims, the same logic applies, but it is most transparent to state burdens of proof directly in terms of coverage, selection/nonresponse, and measurement relative to the stated estimand and target population. Where no defensible source population can be specified, the relevant assumptions concern the link between the realised sample and the target population. Figure 1 is used only to make these burdens of proof explicit; it is not intended to codify a new risk of bias instrument.